# Magnetic Reconfiguration Controls Spin Splitting and Spin-Orbit-Coupled Responses in Compensated Magnets

Yongqian Zhu,[1,*] Lu Liu,[1,*] Xiaobing Chen,[2,†] Tong Zou,[1] Yuntian Liu,[1] Zehao Ji,[1] and Qihang Liu [1,2,†]

[1]*State key laboratory of quantum functional materials, Department of Physics, and Guangdong Basic Research Center of Excellence for Quantum Science, Southern University of Science and Technology (SUSTech), Shenzhen 518055, China*

[2]*Quantum Science Center of Guangdong-Hong Kong-Macao Greater Bay Area (Guangdong), Shenzhen 518045, China*

[*]These authors contributed equally to this work.

[†]Corresponding author: chenxiaobing@quantumsc.cn, liuqh@sustech.edu.cn

*Abstract*—Compensated magnets provide a versatile platform for spin-dependent electronic and transport phenomena. Here, we show that magnetic reconfiguration, *i.e.*, selective reversal of local moments without changing the underlying crystal structure, can profoundly transform the nonrelativistic spin splitting and spin-orbit-coupled responses of spin-degenerate antiferromagnets (AFMs), altermagnets (AMs), and compensated ferrimagnets (cFIMs). Strikingly, magnetic configurations with nearly degenerate energies can exhibit qualitatively different spin-split band structures and anomalous Hall responses, allowing a small magnetic-energy scale to produce large changes in physical properties. Combining tight-binding models, oriented spin space group analysis, and first-principles calculations, we demonstrate that converting an AM configuration into cFIM state activates staggered onsite-potential contributions to spin splitting, complementing the anisotropic-hopping mechanism of AMs and removing symmetry-enforced degeneracies. Moreover, cFIM permits an anomalous Hall response at first order in spin-orbit coupling, in contrast to the higher-order or orientation-constrained responses of corresponding AM states. We demonstrate this regime in $Mn_2Mo_3O_8$, one of 12 cFIM candidates identified in the MAGNDATA database, and further demonstrate its generality by designing cFIMs from collinear AFMs, exemplified by $MnGeO_3$ and $Co_2Mo_3O_8$. Our results establish magnetic reconfiguration as a powerful symmetry-guided strategy for dramatically reshaping electronic and transport responses in compensated magnets.

*Introduction*—Compensated magnets can be broadly divided into two symmetry classes according to whether their vanishing spin magnetization is enforced by symmetry [1]. The first comprises collinear antiferromagnets (AFMs), in which opposite-spin sublattices are connected by spin-group symmetries. In conventional spin-degenerate AFMs, these sublattices are related by inversion or fractional translation, enforcing nonrelativistic spin degeneracy throughout the Brillouin zone [2,3]. In contrast, in altermagnets (AMs), opposite-spin sublattices are connected by spatial rotations, allowing nonrelativistic spin splitting while retaining characteristic symmetry-protected degeneracies [4-22]. Their spin-orbit-coupled responses are likewise constrained by magnetic symmetry: the anomalous Hall effect (AHE) may be forbidden, restricted to particular spin orientations, or allowed only at higher orders in spin-orbit coupling (SOC). Thus, in both spin-degenerate and spin-split AFMs, magnetic symmetry imposes strong constraints on physical responses.

The second class comprises compensated ferrimagnets (cFIMs), in which opposite-spin sublattices are crystallographically inequivalent, despite their vanishing net magnetization [15,23-31]. The absence of sublattice-connecting symmetry permits nonrelativistic spin splitting and anomalous Hall responses. Previous studies have demonstrated the interconversion between AM and cFIM states through chemical substitution [31]. However, it remains unclear how magnetic reconfiguration alone, without modifying the crystal structure, changes the microscopic mechanisms of spin splitting and the leading SOC order of the AHE. A central related question is whether nearly degenerate magnetic configurations can exhibit qualitatively different electronic and transport properties, allowing magnetic symmetry, rather than a large magnetic-energy scale, to govern dramatic changes in physical responses.

Here, we address these questions by combining tight-binding models, oriented spin space group analysis, and first-principles calculations. We show that converting an AM configuration into a cFIM state through selective reversal of local moments activates staggered onsite-potential contributions to spin splitting, complementing the anisotropic-hopping mechanism of AMs and removing symmetry-enforced degeneracies. The reconfiguration also changes the symmetry constraints on spin-orbit-coupled responses: cFIM symmetry permits an anomalous Hall response at first order in SOC, in contrast to the higher-order or orientation-constrained responses of corresponding AM states. Taking $Mn_2Mo_3O_8$ as a representative example among the 12 cFIM candidates identified from MAGNDATA, we demonstrate that the AM and cFIM configurations differ in energy by only 1.0 meV/Mn, yet the latter exhibits strongly enhanced spin splitting and anomalous Hall response. We further identify 46 collinear AFMs from the MAGNDATA database and illustrate the generality of this strategy using $MnGeO_3$ and $Co_2Mo_3O_8$.

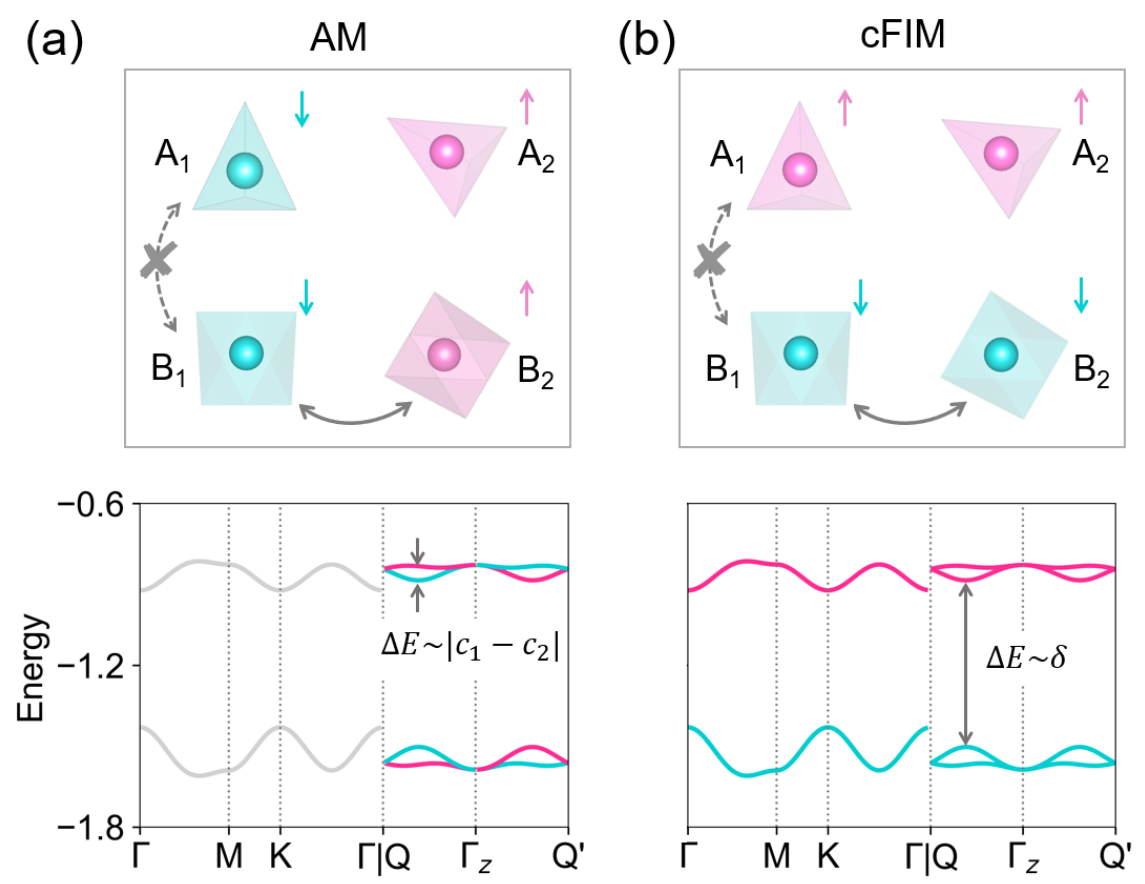

FIG. 1. Schematic illustration of the magnetic configurations and corresponding tight-binding band structures for (a) AM and (b) cFIM phases. The upper panels illustrate the four magnetic sites $A_1$, $A_2$, $B_1$, and $B_2$, where pink and cyan colors denote opposite magnetic moments. The solid bidirectional arrows indicate symmetry connection within each sublattice, whereas the dashed arrows marked by a cross indicate the absence of symmetry connection between the A ($A_1$ and $A_2$) and B ($B_1$ and $B_2$) sublattices. The lower panels show spin-resolved bands calculated from the tight-binding model using parameters $c_1 = 0.015$, $c_2 = 0.005$, $c_1' = -0.01$, $c_2' = -0.002$, $\Delta = 1.55$, and $\delta = 0.7$. Here only four representative bands are shown. The energy splitting $\Delta E$ between opposite spin bands is indicated by vertical arrows. Gray curves indicate spin-degenerate bands, while pink and cyan denote opposite spin channels. The high-symmetry points correspond to Q = (5/12, 1/6, 0.3), $\Gamma_z$ = (0,0,0.3), and Q′ = (−1/6, 7/12, 0.3) in fractional reciprocal coordinates.

*Different mechanisms of spin splitting*—In typical AMs, the momentum-dependent spin splitting is governed by the ligand-mediated anisotropic electron hopping, where nonmagnetic ions lower the environment symmetry of magnetic ions. In insulators, this hopping difference is often suppressed, leading to small electronic spin splitting and magnonic chiral splitting [32,33]. Here, we show that magnetic reconfiguration into a cFIM state overcomes this limitation by activating a staggered onsite potential, which produces giant spin splitting throughout the whole Brillouin zone.

Among the 36 cFIM candidates previously identified through spin-space-group classification [1], our first-principles calculations without spin–orbit coupling (SOC) characterize nine collinear semiconducting cFIMs and three nearly compensated coplanar ferrimagnets after excluding metallic compounds, f-electron-containing materials, and structures with fractional atomic-site occupancies (see Sec. II of the Supplemental Material for details). We select collinear $Mn_2Mo_3O_8$ as a representative example to investigate how magnetic reconfiguration changes the microscopic contributions to nonrelativistic spin splitting. Its primitive cell contains four Mn ions occupying two inequivalent Wyckoff positions: tetrahedrally coordinated A sites ($A_1$ and $A_2$) and octahedrally coordinated B sites ($B_1$ and $B_2$). We first consider an AM phase [Fig. 1(a)], which belongs to the spin space group (SSG) $P^{-1}6_3{}^{1}m^{-1}c^{\infty m}1$ [34-40], where opposite-spin ions in the same crystallographic sublattice are both related by the six-fold screw operation. Reversing the magnetic moments at $A_1$ and $B_2$ produces a cFIM phase with the SSG $P^{1}6_3{}^{1}m^{1}c^{\infty m}1$ [Fig. 1(b)], where the moments are parallel within each sublattice but antiparallel between A and B. Since these sublattices are crystallographically inequivalent, no SSG operation relates their opposite moments. The two phases therefore provide a direct comparison of spin splitting under different SSG symmetries within the same crystal structure.

To capture their essential distinction, we construct an eight-band tight-binding Hamiltonian expressed on the four magnetic sites $i = 1,2,3,4$:

$$H_i(k) = [h_i(k) + \delta_i]\sigma_0 + \Delta_i\sigma_z, \qquad (1)$$

where $\sigma$ represents the Pauli matrices, and $\delta$ denotes the staggered potential taking $\delta_{1,2} = 0$ for A sites and $\delta_{3,4} = \delta$ for B sites. For simplicity, here we retain a minimal form that includes only the diagonal terms sufficient to capture the anisotropic hopping within each sublattice. The site-dependent kinetic terms are $h_i(k) = t_1 \cdot f(k_x, k_y, k_z) + t_2 \cdot f(k_x, k_y, -k_z)$, where the hopping pairs $(t_1, t_2)$ take $(c_1, c_2)$ and $(c_2, c_1)$ for A sites ($i = 1, 2$), and $(c_1', c_2')$ and $(c_2', c_1')$ for B sites ( $i = 3, 4$ ). Here, $f(k_x, k_y, k_z) = 2\cos(\sqrt{3}k_y - k_z) + 2\cos\left(\frac{3k_x}{2} + \frac{\sqrt{3}k_y}{2} + k_z\right) + 2\cos\left(\frac{3k_x}{2} - \frac{\sqrt{3}k_y}{2} - k_z\right)$. The magnetic configurations are specified by the local exchange term $\Delta$ by taking $(-\Delta, -\Delta, +\Delta, +\Delta)$ for cFIM and $(+\Delta, -\Delta, +\Delta, -\Delta)$ for AM.

The resulting band structures reveal distinct microscopic origins of spin splitting, as shown in lower panels of Fig. 1. In the AM phase, spin splitting emerges only in restricted momentum regions and vanishes along symmetry-protected nodal lines. It is governed by anisotropic electron hopping and scales with $|c_1 - c_2|$. In insulators where inter-site electron hopping decays rapidly, this non-local spatial anisotropy is heavily suppressed to around 0.01~0.1 eV. In contrast, spin splitting in the cFIM phase is directly driven by the staggered onsite potential $\delta$, which originates from the distinct local crystal-field environments of sites A and B. As a local term insensitive to spatial decay, $\delta$ provides a dominant energy scale around 0.1~1 eV. Consequently, the cFIM phase shows giant spin splitting throughout the entire Brillouin zone. Although illustrated using $Mn_2Mo_3O_8$, this mechanism can extend to other AM–cFIM configurations with inequivalent magnetic sites.

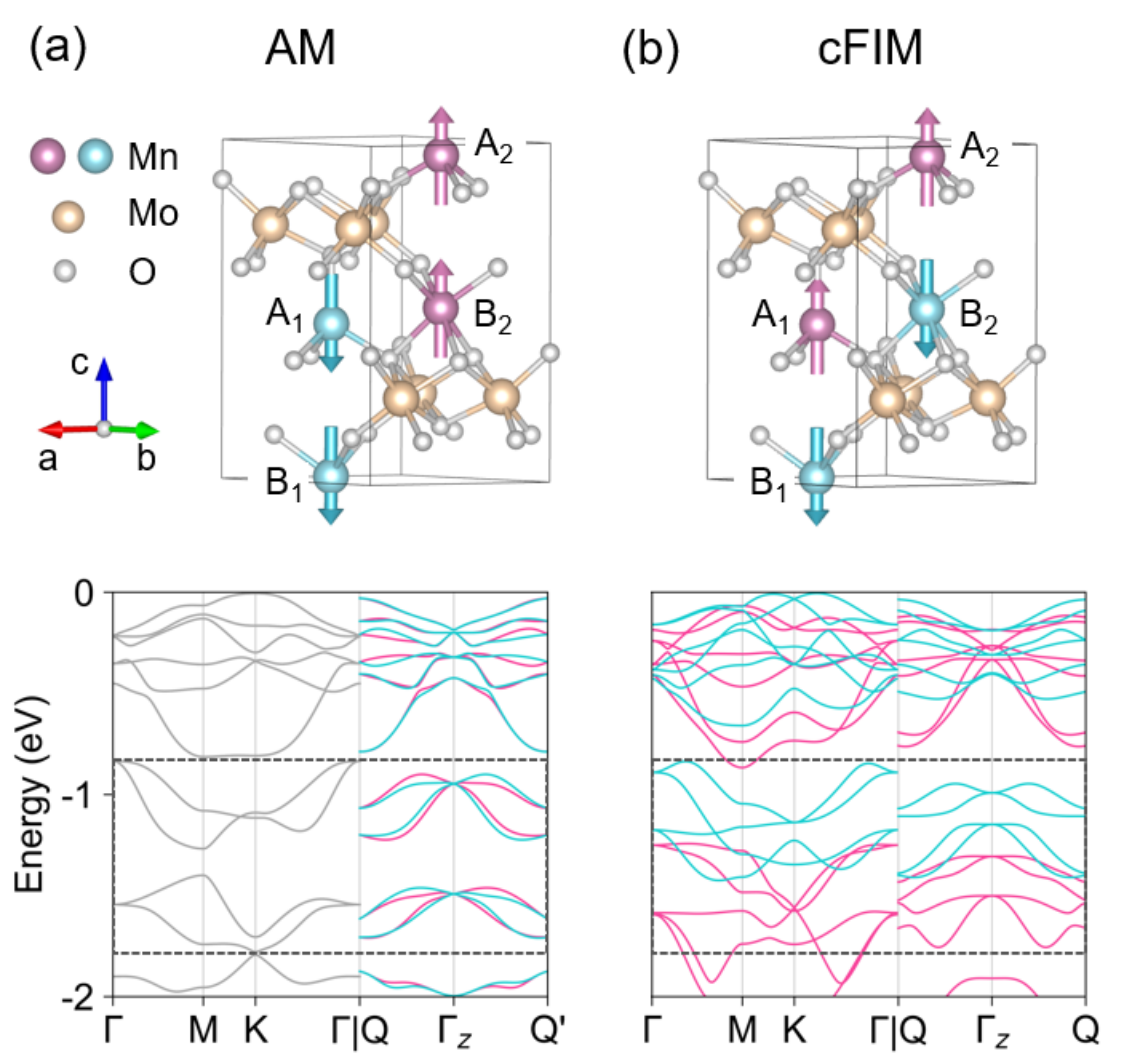

FIG. 2. Crystal structures and first-principles band structures of $Mn_2Mo_3O_8$ in the (a) AM and (b) cFIM phases. In upper panels, four Mn atoms are labeled as $A_1$, $A_2$, $B_1$ and $B_2$, where pink and cyan arrows denote opposite magnetic moments. The lower panels show spin-resolved band structures calculated without SOC. Gray curves denote spin-degenerate bands, while pink and cyan indicate opposite spin channels. The dashed rectangle highlights bands with substantial Mn-O hybridization, for which the Mn contribution is mainly from the $d_{xz}$ and $d_{yz}$ orbitals. The Fermi energy is set to zero.

*Low-energy magnetic reconfiguration in $Mn_2Mo_3O_8$*—To verify the tight-binding model, we perform first-principles calculations for $Mn_2Mo_3O_8$ [41,42] in both AM and cFIM configurations. The two magnetic states exhibit a sharp contrast: while their spin-resolved electronic structures differ markedly, their total energies are nearly degenerate, with the AM state lying merely 1.0 meV/Mn higher than the cFIM ground state. This small energy difference can be captured by an effective Heisenberg spin model [43],

$$H = -\sum_{ij} J_{ij} S_i \cdot S_j \,, \qquad (2)$$

where the dominant nearest-neighbor exchange interaction $J_{A_1B_2} = -5.5$ meV contributes identically to both configurations, canceling out in the energy difference between two states. Consequently, the energy difference is governed by weaker further-neighbor interactions (below 0.7 meV in magnitude), whose accumulated contributions largely cancel owing to their competing signs. The resulting energy difference of 0.84 meV/Mn agrees reasonably well with the first-principles result (see Supplemental Material for details).

The calculated band structures confirm the symmetry-breaking mechanism (lower panels of Fig. 2). In the AM phase, the sublattice-connecting symmetry $\{\bar{1}||2_{001}\}$ enforces the spin degeneracy along the Γ–M–K–Γ path. Away from these nodal lines, the spin splitting remains weak. The bands highlighted by the dashed rectangle exhibit substantial Mn-O hybridization, with the Mn contribution arising mainly from the $d_{xz}$ and $d_{yz}$ orbitals. Within this orbital sector, symmetry analysis identifies fifth-neighbor hopping as the lowest-order contribution capable of inducing AM spin splitting, with the corresponding Wannier hopping matrix elements below 2 meV. In contrast, in the cFIM phase, the $\{\bar{1}||2_{001}\}$ operation is broken, directly activating the staggered onsite potential of 0.42 eV in the same Wannier orbital sector. As a result, giant spin splitting emerges in the whole Brillouin zone. Thus, nearly degenerate magnetic configurations can exhibit markedly different spin-split electronic structures.

*First-order SOC dependence of anomalous Hall effect*—Beyond modifying spin splitting, the magnetic reconfiguration from AM to cFIM renders the anomalous Hall effect independent of magnetization orientation while ensuring that the transport response emerges at first order in SOC. The anomalous Hall conductivity (AHC) pseudovector $\boldsymbol{\sigma}^{\mathrm{AHC}} = (\sigma_{yz}, \sigma_{zx}, \sigma_{xy})$ transforms identically to the orbital magnetization, which is strictly forbidden by the spin mirror symmetry in collinear magnets in the absence of SOC [44,45]. While the magnetic space group (MSG) determines whether an AHC component is allowed [46-49], the oriented spin space group (OSSG) further constrains its leading SOC order [1,50,51]. To see this, we write the SOC Hamiltonian as $\hat{H}_{SOC} = \lambda \hat{\boldsymbol{L}}^T \boldsymbol{\chi} \hat{\boldsymbol{\sigma}} = \lambda \sum_{ij} \chi_{ij} \hat{L}_i \hat{\sigma}_j$ , where λ, $\boldsymbol{\chi}$, $\hat{\boldsymbol{L}}$ and $\hat{\boldsymbol{\sigma}}$ denote the SOC strength, SOC tensor, effective orbital angular momentum operator, and spin operator, respectively. In the weak-SOC regime, the spin and orbital magnetization, $\boldsymbol{M}_s$ and $\boldsymbol{M}_o$, can be expanded perturbatively in powers of $\lambda\boldsymbol{\chi}$. A single contraction with the SOC tensor connects their symmetry transformation properties, namely $\boldsymbol{\chi}\boldsymbol{M}_s \sim \boldsymbol{M}_o$ and $\boldsymbol{\chi}^T \boldsymbol{M}_o \sim \boldsymbol{M}_s$ , where ~ denotes transformation according to the same symmetry representation. Consequently, their lowest

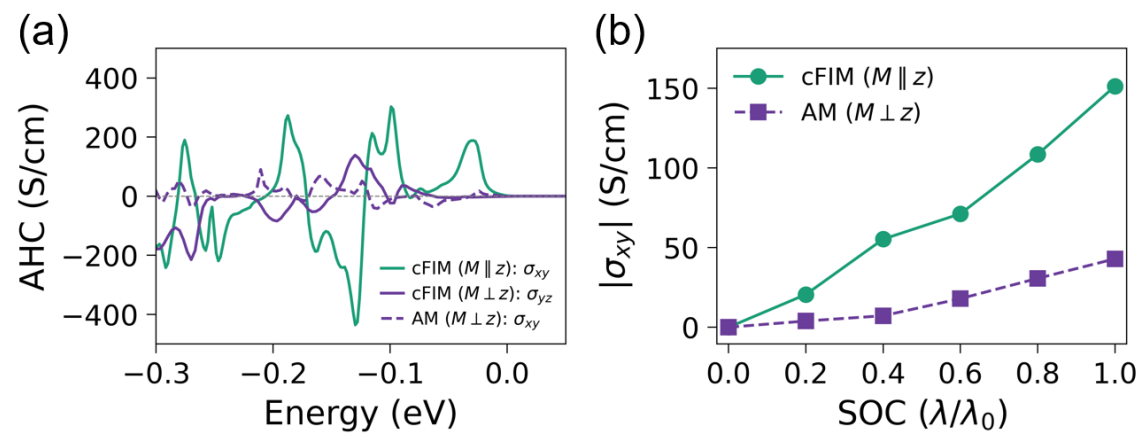

FIG. 3. (a) AHC calculated with SOC included. Here, the $x$ and $z$ axes are taken along the crystalline $a$ and $c$ directions, respectively. $M \parallel z$ and $M \perp z$ denote magnetic moments along $z$ and in the $xy$ plane at 30° from $x$, respectively. The valence-band maximum is set to zero. (b) Dependence of the magnitude of $\sigma_{xy}$ on normalized SOC strength $\lambda/\lambda_0$ at the energy of $-0.11$ eV, where $\lambda_0$ denotes the actual SOC strength.

symmetry-allowed SOC orders differ by at most one [52].

In the AM phase, the SSG forbids zeroth-order spin magnetization, whether AHC is allowed depends on the magnetization orientation that dictates the corresponding MSG; when allowed, its leading SOC order is governed by the symmetry connecting opposite-spin sublattices [53]. In contrast, the SSG of a collinear cFIM permits zeroth-order spin magnetization, even though its actual value vanishes through compensation. This symmetry allowance ensures that at least one AHC component is permitted to emerge at first order in SOC, irrespective of magnetization direction or crystal symmetry.

The theoretical conclusions above are further validated in $Mn_2Mo_3O_8$ via symmetry analysis and DFT calculations. When the magnetic moment aligns along the $z$ direction ($M \parallel z$), the AM phase belongs to the MSG $P6_3'm'c$, which forbids AHC, whereas the cFIM phase adopts MSG $P6_3m'c'$, which allows net magnetization along z and the AHC component $\sigma_{xy}$ [Fig. 3(a)]. We further rotate the magnetic moment into the $xy$ plane at 30° with respect to the $x$ axis ($M \perp z$). This in-plane magnetic moment further lowers the MSGs, inducing $\sigma_{xy}$ in the AM phase while supporting $\sigma_{yz}$ and $\sigma_{zx}$ in the cFIM phase [Fig. 3(a)]. The OSSGs of the cFIM ($M \parallel z$) and AM ($M \perp z$) phases are $P^{1}6_3{}^{1}m^{1}c^{\infty_{001}m}1$ and $P^{-1}6_3{}^{1}m^{-1}c^{\infty_{210}m}1$, respectively. Further SOC tensor expansion demonstrates leading SOC orders of first and third for the AHC in the cFIM and AM phases, respectively. As shown in Fig. 3(b), the calculated AHC in the cFIM grows linearly with SOC strength, whereas that in the AM phase exhibits a third-order dependence with a substantially smaller magnitude. These results illustrate how magnetic reconfiguration into a cFIM activates first-order SOC contributions and enhances the AHC.

*Designing cFIMs from collinear AFMs*—The preceding magnetic reconfiguration suggests a symmetry-guided strategy for designing cFIM states from collinear AFMs, including both conventional AFMs and AMs. Here we focus on collinear AFMs with magnetic atoms occupying multiple inequivalent Wyckoff positions, although the design principle can be extended to more general magnetic configurations, as discussed below. By screening the MAGNDATA database [54,55], we identify 46 candidate materials possessing multiple inequivalent magnetic sublattices with equal local moment magnitudes, as listed in Table S3 of Supplemental Material.

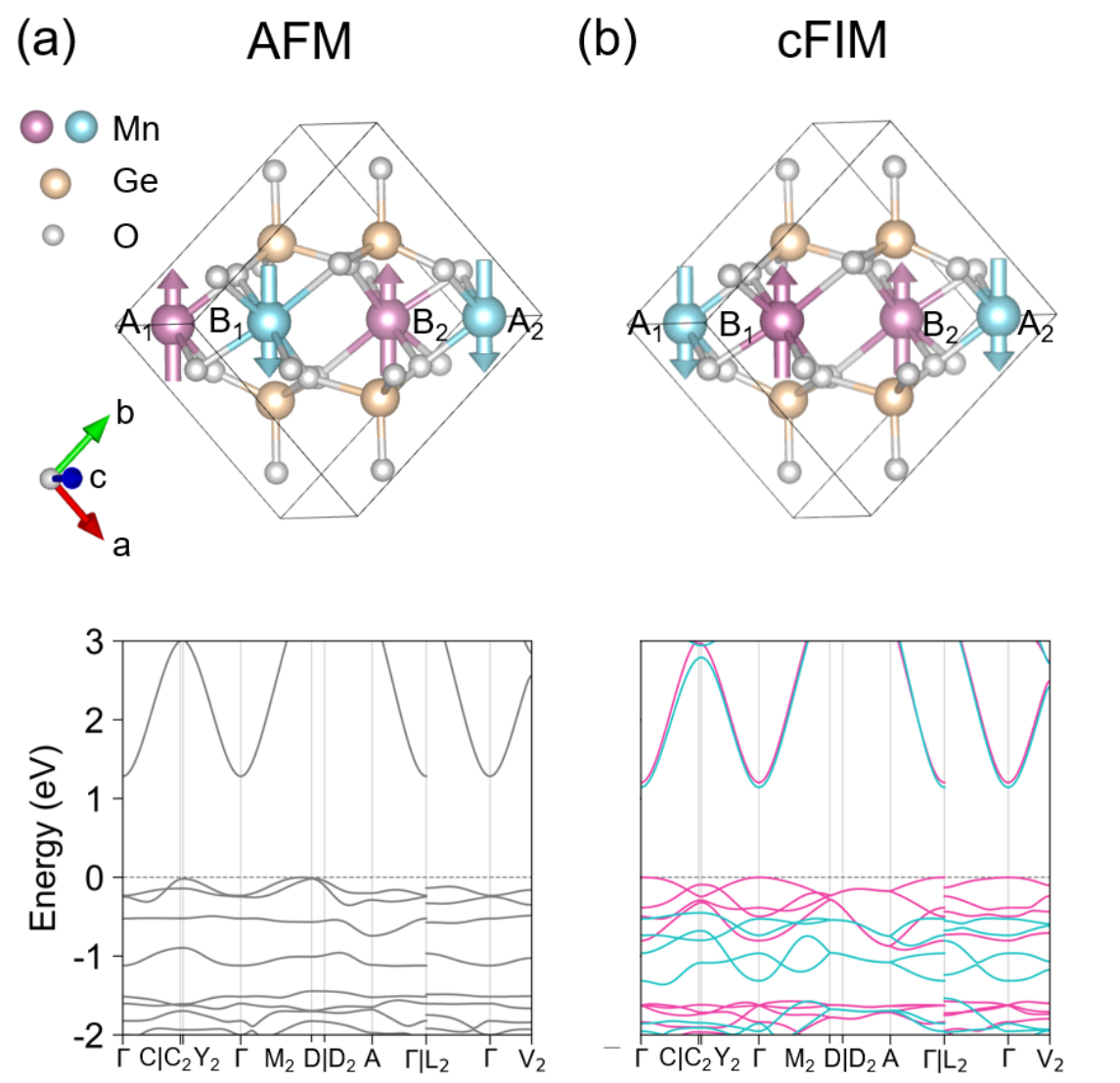

FIG. 4. Crystal structures and first-principles band structures of $MnGeO_3$ in the (a) AFM and (b) cFIM phases. In upper panels, four Mn atoms are labeled as $A_1$, $A_2$, $B_1$ and $B_2$, where pink and cyan arrows denote opposite magnetic moments. The lower panels show spin-resolved band structures calculated without SOC. Gray curves denote spin-degenerate bands, while pink and cyan indicate opposite spin channels. The Fermi energy is set to zero.

Starting from such AFMs, we construct cFIM phases by selectively reversing local moments at different Wyckoff positions, as illustrated in Fig. 1. This operation breaks the symmetry that connects opposite-spin sublattices, thereby lifting the symmetry-enforced compensation. Since the magnetic moments on selected Wyckoff sites have equal magnitudes in the AFM phase, the total magnetization remains zero even after symmetry breaking. This condition is naturally satisfied in insulating or half-metallic systems, where the integer electronic occupation enforced by the band gap stabilize the local magnetic moments according to the Luttinger theorem. In this way, a symmetry-enforced AFM is converted into a filling-enforced cFIM.

The accessibility of such cFIM phases is governed by their energy proximity to the parent AFM phase, enabling magnetic switch under moderate external perturbations, such as magnetic fields, strain, or doping. As a representative example, conventional AFM $MnGeO_3$ [56] adopts the SSG $C^{1}2/^{-1}c^{\infty m}1$ [Fig. 4(a)], where spatial inversion symmetry connects opposite-spin sublattices, enforces spin degeneracy and forbids the AHE. Our calculations show that selectively reversing moments on $A_1$ and $B_1$ sites yield a cFIM phase lying only 2.0 meV/Mn above the AFM phase. This symmetry breaking unlocks pronounced spin splitting throughout the Brillouin zone driven by sublattice-staggered onsite potentials. Moreover, the AHE is forbidden in the AFM phase by the combined inversion and time-reversal symmetry, regardless of the magnetization direction, whereas it becomes symmetry-allowed in the cFIM phase. Similarly, in $Co_2Mo_3O_8$ [41], an altermagnet that shares the same crystal structure as $Mn_2Mo_3O_8$ but has an AM ground state, which is only 1.7 meV/Co lower in energy than cFIM phase (Fig. S13). By selectively reversing the magnetic moments, the AM phase can be converted into the cFIM phase, giving rise to giant spin splitting throughout the Brillouin zone. These results demonstrate that the proposed design strategy provides an effective route to realizing large spin splitting and symmetry-enabled AHE, offering a promising avenue for experimental detection and potential applications.

*Discussion*—The magnetic-reconfiguration strategy has so far focused on collinear compensated magnets with magnetic atoms occupying multiple crystallographic Wyckoff positions. However, it can in principle be extended to systems with magnetic atoms occupying a single Wyckoff position. Two cases can be distinguished. First, the crystallographic Wyckoff position may split into multiple inequivalent spin Wyckoff positions under the SSG. Similar to the case in Fig. 1, cFIM configurations can then be constructed from these AFMs by selectively reversing local moments at different spin Wyckoff positions to break the symmetry connecting opposite-spin sublattices while preserving magnetization compensation. Second, all magnetic atoms may initially belong to a single spin Wyckoff position under the SSG. In this case, selective reversal of a suitable subset of local moments can lower the SSG symmetry, splitting the original spin Wyckoff position into inequivalent ones while maintaining zero net spin magnetization. For equal-magnitude moments, this requires balanced reversals of oppositely oriented moments, potentially involving an enlarged magnetic unit cell. Thus, multiple crystallographic Wyckoff positions are not a necessary condition for constructing cFIM configurations.

Beyond collinear systems, the magnetic reconfiguration can also extend to nearly compensated noncollinear ferrimagnets. For example, $Ni_3B_7O_{13}Cl$ [57] consists of two orthogonal collinear spin subsystems (Table S1 and Fig. S10 of the Supplemental Material). One subsystem, associated with a single spin Wyckoff position, is symmetry-compensated, whereas the other occupies two crystallographic Wyckoff positions with nearly equal but oppositely oriented moments, resulting in near compensation with $0.0038\ \mu_B$/Ni. The latter resembles the collinear cFIM configuration of $Mn_2Mo_3O_8$. Selective reversal of local moments can transform this nearly compensated ferrimagnetic configuration into a noncollinear AFM configuration. Conversely, the inverse operation provides a route from the AFM configuration to the nearly compensated ferrimagnetic state, illustrating an analogous correspondence to the collinear case. More generally, nearly compensated ferrimagnetic configurations can be constructed from noncollinear AFMs by selectively reversing a subset of local moments whose vector sum approximately vanishes, while breaking the symmetry that originally enforced magnetization compensation.

*Conclusions*—In summary, we establish how magnetic reconfiguration can profoundly alter spin-dependent electronic and transport responses in compensated magnets. The nearly degenerate AM and cFIM configurations of $Mn_2Mo_3O_8$ demonstrate that a small magnetic energy difference can coexist with markedly different spin splittings, while the associated symmetry breaking also modifies the leading SOC order of the AHE. More broadly, our findings establish the design of cFIMs from AFMs as a symmetry-guided route to engineering electronic and transport properties, opening new opportunities for high-performance spintronic applications.

*Acknowledgement*—This work was supported by National Natural Science Foundation of China under Grant Nos. 12525410, 12274194, 12574275, and 12534003, Guangdong Provincial Quantum Science Strategic Initiative under Grant No. GDZX2401002, Guangdong Provincial Key Laboratory for Computational Science and Material Design under Grant No. 2019B030301001, Shenzhen Science and Technology Program (Grant Nos. RCJC20221008092722009, 20231117091158001 and QNXMC20250701094702004), the Innovative Team of General Higher Educational Institutes in Guangdong Province (Grant No. 2020KCXTD001), and Center for Computational Science and Engineering of Southern University of Science and Technology.